\PassOptionsToPackage{unicode}{hyperref}
\PassOptionsToPackage{hyphens}{url}
\PassOptionsToPackage{dvipsnames,svgnames,x11names}{xcolor}
\documentclass[
  12pt]{article}

\usepackage{amsmath,amssymb,amsthm}
\usepackage{iftex}
\ifPDFTeX
  \usepackage[T1]{fontenc}
  \usepackage[utf8]{inputenc}
  \usepackage{textcomp} 
\else 
  \usepackage{unicode-math}
  \defaultfontfeatures{Scale=MatchLowercase}
  \defaultfontfeatures[\rmfamily]{Ligatures=TeX,Scale=1}
\fi
\usepackage{lmodern}
\ifPDFTeX\else
\fi
\IfFileExists{upquote.sty}{\usepackage{upquote}}{}
\IfFileExists{microtype.sty}{
  \usepackage[]{microtype}
  \UseMicrotypeSet[protrusion]{basicmath} 
}{}
\makeatletter
\@ifundefined{KOMAClassName}{
  \IfFileExists{parskip.sty}{%
    \usepackage{parskip}
  }{
    \setlength{\parindent}{0pt}
    \setlength{\parskip}{6pt plus 2pt minus 1pt}}
}{
  \KOMAoptions{parskip=half}}
\makeatother
\usepackage{xcolor}
\makeatletter
\ifx\paragraph\undefined\else
  \let\oldparagraph\paragraph
  \renewcommand{\paragraph}{
    \@ifstar
      \xxxParagraphStar
      \xxxParagraphNoStar
  }
  \newcommand{\xxxParagraphStar}[1]{\oldparagraph*{#1}\mbox{}}
  \newcommand{\xxxParagraphNoStar}[1]{\oldparagraph{#1}\mbox{}}
\fi
\ifx\subparagraph\undefined\else
  \let\oldsubparagraph\subparagraph
  \renewcommand{\subparagraph}{
    \@ifstar
      \xxxSubParagraphStar
      \xxxSubParagraphNoStar
  }
  \newcommand{\xxxSubParagraphStar}[1]{\oldsubparagraph*{#1}\mbox{}}
  \newcommand{\xxxSubParagraphNoStar}[1]{\oldsubparagraph{#1}\mbox{}}
\fi
\makeatother

\usepackage{longtable,booktabs,array}
\usepackage{calc} 
\usepackage{etoolbox}
\makeatletter
\patchcmd\longtable{\par}{\if@noskipsec\mbox{}\fi\par}{}{}
\makeatother
\IfFileExists{footnotehyper.sty}{\usepackage{footnotehyper}}{\usepackage{footnote}}
\makesavenoteenv{longtable}
\usepackage{graphicx}
\makeatletter
\def\maxwidth{\ifdim\Gin@nat@width>\linewidth\linewidth\else\Gin@nat@width\fi}
\def\maxheight{\ifdim\Gin@nat@height>\textheight\textheight\else\Gin@nat@height\fi}
\makeatother
\setkeys{Gin}{width=\maxwidth,height=\maxheight,keepaspectratio}
\makeatletter
\def\fps@figure{htbp}
\makeatother

\makeatletter
\@ifpackageloaded{caption}{}{\usepackage{caption}}
\AtBeginDocument{%
\ifdefined\contentsname
  \renewcommand*\contentsname{Table of contents}
\else
  \newcommand\contentsname{Table of contents}
\fi
\ifdefined\listfigurename
  \renewcommand*\listfigurename{List of Figures}
\else
  \newcommand\listfigurename{List of Figures}
\fi
\ifdefined\listtablename
  \renewcommand*\listtablename{List of Tables}
\else
  \newcommand\listtablename{List of Tables}
\fi
\ifdefined\figurename
  \renewcommand*\figurename{Figure}
\else
  \newcommand\figurename{Figure}
\fi
\ifdefined\tablename
  \renewcommand*\tablename{Table}
\else
  \newcommand\tablename{Table}
\fi
}
\@ifpackageloaded{float}{}{\usepackage{float}}
\floatstyle{ruled}
\@ifundefined{c@chapter}{\newfloat{codelisting}{h}{lop}}{\newfloat{codelisting}{h}{lop}[chapter]}
\floatname{codelisting}{Listing}

\makeatother
\makeatletter
\@ifpackageloaded{caption}{}{\usepackage{caption}}
\@ifpackageloaded{subcaption}{}{\usepackage{subcaption}}
\makeatother

\ifLuaTeX
  \usepackage{selnolig}  
\fi
\usepackage[]{natbib}
\usepackage{bookmark}

\IfFileExists{xurl.sty}{\usepackage{xurl}}{} 
\hypersetup{
  pdftitle={Retrospective Orthogonal Design: Response-Surface Reconstruction from Observational Data},
  pdfauthor={Lawrence V. Fulton; Christopher P. Fulton; Arvind Sharma; Aleksandar Tomi\'{c}},
  pdfkeywords={response-surface reconstruction; observational data; retrospective design diagnostics; variance decomposition; simulation sensitivity analysis},
  colorlinks=true,
  linkcolor={blue},
  filecolor={Maroon},
  citecolor={Blue},
  urlcolor={Blue},
  pdfcreator={LaTeX via pandoc}}

\usepackage{multirow}
\usepackage{makecell}
\usepackage{colortbl}
\usepackage{adjustbox}
\usepackage{tikz}
\usepackage{pgfplots}
\pgfplotsset{compat=1.18}
\usetikzlibrary{
    shapes.geometric,
    arrows.meta,
    positioning,
    fit,
    backgrounds,
    decorations.pathreplacing,
    calc
}
\graphicspath{{fig/}}

\newtheorem{theorem}{Theorem}
\newtheorem{corollary}{Corollary}
\newtheorem{proposition}{Proposition}

\newcommand{\anon}{1}

\begin{document}

\def\spacingset#1{\renewcommand{\baselinestretch}%
{#1}\small\normalsize} \spacingset{1}


\if1\anon
{
  \title{\bf Retrospective Orthogonal Design: Response-Surface Reconstruction from Observational Data}
  \author{Lawrence V. Fulton\thanks{
    Corresponding author. This research received no external funding.}\hspace{.2cm}\\
    Applied Analytics, Boston College\\
    and \\
    Christopher P. Fulton \\
    United States Air Force Test Pilot School, Edwards Air Force Base\\
    and \\
    Arvind Sharma \\
    Applied Analytics, Boston College\\
    and \\
    Aleksandar Tomi\'{c} \\
    Applied Analytics, Boston College}
  \maketitle
} \fi

\if0\anon
{
  \bigskip
  \bigskip
  \bigskip
  \begin{center}
    {\LARGE\bf Retrospective Orthogonal Design: Response-Surface Reconstruction from Observational Data}
\end{center}
  \medskip
} \fi

\bigskip
\begin{abstract}
Regression estimates from observational data can depend on specification under multicollinearity, while sequential sums of squares (SS) depend on term order. We introduce Retrospective Orthogonal Design (ROD), which reconstructs conditional mean surfaces on a probability-balanced lattice. ROD preserves observed cell means, completes unsupported cells, applies weighted tensor-product contrasts, and evaluates the reconstructed surface through piecewise-affine interpolation over Freudenthal polyhedra. Resolution and completion are selected jointly by validation among rank-admissible candidates, followed by refitting and evaluation on an untouched test set. For an admissible lattice, $\mathbf{X}^{\top}\mathbf{W}\mathbf{X}=c\mathbf{I}$, yielding specification-invariant contrast effects and unique, order-independent SS within the retained contrast space. Response-free projection calibration maps the fixed reconstruction onto a declared scientific basis and corrects finite-resolution recovery loss. Across 6,480 simulation conditions spanning nine data-generating processes, ROD matched or exceeded polynomial regression in five processes and performed strongest on threshold, sign-interaction, and localized surfaces. For the quadratic-interaction process, mean out-of-sample $R^2$ differed by only $0.0001$, while calibrated coefficient bias remained small across prespecified targets. A Rao-based information adjustment provides dependence-aware sample-size guidance for ROD planning. In a weighted Mincer application, ROD produced the highest out-of-sample $R^2$ point estimate, with substantial interval overlap with polynomial regression, and provided exhaustive SS allocations invariant to term-entry order.
\end{abstract}

\noindent%
\vfill

\newpage
\spacingset{1.8} 

\section{Introduction}\label{sec-intro}

Regression and ANOVA are special cases of the general linear model, \(\mathbf{y}=\mathbf{X}\boldsymbol{\beta}+\boldsymbol{\varepsilon}\), but differ in how their design matrices are constructed. In a balanced experimental design with a full-rank orthogonal contrast basis, \(\mathbf{X}^{\top}\mathbf{X}=\mathbf{D}\), diagonal, or \(\mathbf{X}^{\top}\mathbf{X}=c\mathbf{I}\) under common-norm scaling \citep{montgomery2019design}. Orthogonality yields additive, order-invariant sums of squares (SS), and retained coefficients are unchanged when additional orthogonal terms enter. Under independent homoskedastic errors and a fixed design, orthogonal estimators are uncorrelated and Gaussian estimators independent, although sampling independence, unlike algebraic orthogonality, still depends on the reconstruction-error covariance \citep{WuHamada2021,Christensen2020,LinStufken2025}. Observational studies rarely possess this structure: correlated predictors make \(\mathbf{X}^{\top}\mathbf{X}\) non-diagonal, so attribution, unique variance, and precision depend on the retained specification and predictor distribution \citep{Belsley1980}. Adding controls or choosing among defensible specifications can materially alter estimated effects \citep{LenzSahn2021,Ratkovic2023,GirardiEtAl2024}, while polynomial and interaction expansions induce further dependence among lower-order terms, products, and powers \citep{WurmReitan2025}.

Existing remedies address parts of this problem without recovering an interpretable orthogonal response surface. Principal components regression (PCR) rotates correlated predictors into orthogonal components, gaining stability at the cost of interpretation in the original variables \citep{ChanEtAl2022}. Orthogonal polynomial contrasts separate trends within a predictor but do not remove cross-predictor dependence \citep{Kutner2005,RaynerLivingston2024,GranziolEtAl2025}. Centering and scaling do not remove substantive collinearity \citep{WurmReitan2025}, while coarsening methods target covariate balance rather than orthogonal surface reconstruction \citep{GhoshWang2026}. Nonparametric alternatives face related tradeoffs: regressograms and multivariate histogram regression estimate conditional means over partitions \citep{Nobel1996,GyorfiEtAl2002}, regression trees use response-adaptive partitions \citep{BreimanEtAl1984}, orthogonal-series estimators approximate the conditional mean in a finite or expanding basis \citep{Newey1997}, and GAMs or tensor-product splines estimate regularized smooth components \citep{HastieTibshirani1986,Wood2006}. These methods can flexibly represent nonlinear structure, but their global attribution remains dependent on partition, basis, knots, smoothing, or the empirical predictor distribution, and exact empirical orthogonality is generally not preserved.

Hoeffding decompositions, functional ANOVA, and Sobol methods provide the closest theoretical precedent for orthogonal variance attribution by decomposing a response function under a specified input measure \citep{Hoeffding1948,Hooker2007,SaltelliEtAl2010}. Classical formulations typically use the product measure \(d\mu(\mathbf{x})=\prod_{j=1}^{k}d\mu_j(x_j)\), under which component functions are mutually orthogonal and variance decomposes additively. Predictor dependence breaks this construction: the analyst must either impose an independent reference measure, so attribution refers to a declared distribution rather than the observed joint distribution, or adopt a dependent-input generalization that sacrifices exact orthogonality. Propensity-score weighting, entropy balancing, and stable balancing similarly alter the predictor distribution, but for causal or missing-data estimands rather than conditional-mean reconstruction or orthogonal variance decomposition \citep{RosenbaumRubin1983,Hainmueller2012,ImaiRatkovic2014,Zubizarreta2015}. The distinction matters because weighting for covariate balance and weighting for response-surface reconstruction serve different estimands.

Classical regression represents \(f(\mathbf{x})\) through coefficients attached to a chosen basis \citep{Myers2016}. Retrospective Orthogonal Design (ROD) instead approximates \(f(\mathbf{x})\) on a probability-balanced tessellation and treats coefficients, marginal effects, and variance decompositions as projections of the reconstructed surface. ROD combines probability-scale discretization, orthonormal contrast coding, normalization weighting, cell completion, and piecewise-affine evaluation to construct an admissible balanced cell-mean lattice. Although each component has precedent, their integration yields exact contrast orthogonality, cell-level provenance, unique conditional SS attribution, continuous evaluation, and response-free recovery calibration within one reconstruction framework.

Like product-measure decompositions, ROD uses an independent reference measure to secure exact orthogonality under dependent predictors, but makes it explicit and fixed rather than allowing attribution to vary with observed covariate density. For an admissible reconstruction, \(\mathbf{X}^{\top}\mathbf{W}\mathbf{X}=c\mathbf{I}\), yielding exact algebraic orthogonality, specification-invariant contrast effects, and order-independent SS within the retained contrast space. Sampling independence remains conditional on reconstruction-error covariance. When a structural basis is declared, response-free calibration maps the reconstructed surface to those coordinates without changing its native orthogonal decomposition. An operational Python implementation provides reconstruction, completion, validation, interpolation, calibration, projection, and diagnostic procedures developed here, with software and source code available at \url{https://github.com/dustoff06/ROD}.

ROD is not derived from an observation-level Gaussian residual model and does not orthogonalize the observed predictors. Its estimand is the reconstructed conditional mean surface under the declared reference measure, and its finite-sample behavior depends on support, completion, conditioning, orthogonality, and provenance. Native contrasts and SS characterize that reconstruction, while calibrated projections provide structural summaries when a target basis is declared. The simulations therefore evaluate both predictive and structural recovery while tracking completion burden, rank admissibility, and reconstruction performance.

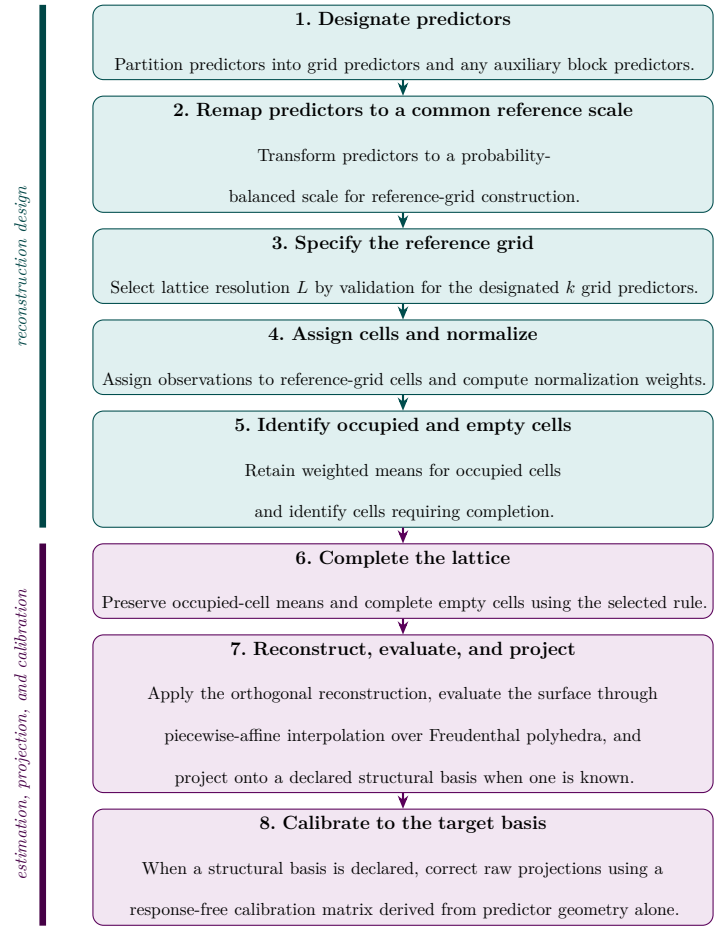
\begin{figure}[H]
\centering
\begin{tikzpicture}[
scale=0.6,
transform shape,
node distance=0.35cm,
tealbox/.style={
rectangle,
rounded corners=4pt,
minimum width=13.6cm,
minimum height=1.55cm,
align=center,
text width=13.4cm,
draw=teal!60!black,
line width=0.4pt,
fill=teal!12
},
purplebox/.style={
rectangle,
rounded corners=4pt,
minimum width=13.6cm,
minimum height=1.85cm,
align=center,
text width=13.4cm,
draw=violet!70!black,
line width=0.4pt,
fill=violet!10
},
arr/.style={
-{Stealth[length=5pt,width=4pt]},
line width=0.8pt
},
]

\node[tealbox] (S1) {%
\textbf{1.\ Designate predictors}\\[2pt]
{\small Partition predictors into grid predictors and any auxiliary block predictors.}};

\node[tealbox, below=of S1] (S2) {%
\textbf{2.\ Remap predictors to a common reference scale}\\[2pt]
{\small Transform predictors to a probability-balanced scale for reference-grid construction.}};

\node[tealbox, below=of S2] (S3) {%
\textbf{3.\ Specify the reference grid}\\[2pt]
{\small Select lattice resolution \(L\) by validation for the designated \(k\) grid predictors.}};

\node[tealbox, below=of S3] (S4) {%
\textbf{4.\ Assign cells and normalize}\\[2pt]
{\small Assign observations to reference-grid cells and compute normalization weights.}};

\node[tealbox, below=of S4] (S5) {%
\textbf{5.\ Identify occupied and empty cells}\\[2pt]
{\small Retain weighted means for occupied cells and identify cells requiring completion.}};

\draw[dashed, gray!50, line width=0.6pt]
($(S5.south west) + (0,-0.52)$) --
($(S5.south east) + (0,-0.52)$);

\node[purplebox, below=of S5, minimum height=1.55cm] (S6) {%
\textbf{6.\ Complete the lattice}\\[2pt]
{\small Preserve occupied-cell means and complete empty cells using the selected rule.}};

\node[purplebox, below=of S6, minimum height=2.1cm] (S7) {%
\textbf{7.\ Reconstruct, evaluate, and project}\\[2pt]
{\small Apply the orthogonal reconstruction, evaluate the surface through
piecewise-affine interpolation over Freudenthal polyhedra, and project onto a
declared structural basis when one is known.}};

\node[purplebox, below=of S7, minimum height=1.85cm] (S8) {%
\textbf{8.\ Calibrate to the target basis}\\[2pt]
{\small When a structural basis is declared, correct raw projections using a
response-free calibration matrix derived from predictor geometry alone.}};

\draw[arr, teal!60!black] (S1.south) -- (S2.north);
\draw[arr, teal!60!black] (S2.south) -- (S3.north);
\draw[arr, teal!60!black] (S3.south) -- (S4.north);
\draw[arr, teal!60!black] (S4.south) -- (S5.north);

\draw[arr, violet!70!black] (S5.south) -- (S6.north);
\draw[arr, violet!70!black] (S6.south) -- (S7.north);
\draw[arr, violet!70!black] (S7.south) -- (S8.north);

\draw[line width=2.5pt, violet!55!black]
($(S6.north west) + (-1.1,0)$) --
($(S8.south west) + (-1.1,0)$);

\node[
rotate=90,
violet!55!black,
font=\small\itshape,
anchor=center
]
at ($(S6.north west)!0.5!(S8.south west) + (-1.55,0)$)
{estimation, projection, and calibration};

\draw[line width=2.5pt, teal!55!black]
($(S1.north west) + (-1.1,0)$) --
($(S5.south west) + (-1.1,0)$);

\node[
rotate=90,
teal!55!black,
font=\small\itshape,
anchor=center
]
at ($(S1.north west)!0.5!(S5.south west) + (-1.55,0)$)
{reconstruction design};

\end{tikzpicture}
\caption{\label{fig-first}Overview of the ROD workflow.}
\end{figure}

\section{ROD Model}\label{sec-meth}

Figure~\ref{fig-first} summarizes the ROD workflow. Steps~1--5 construct the probability-balanced reference lattice and identify occupied and empty cells, while Steps~6--8 complete, reconstruct, evaluate, project, and calibrate the surface.

In Steps~1--2, predictors are partitioned into grid predictors, whose discretized levels define the factorial contrast space, and block predictors, which enter through standard indicator or effect coding without participating in lattice construction. Ordinal predictors may be treated as grid predictors when their ordering is meaningful. Each grid predictor is mapped to a probability scale before discretization. Let $r_j=F_j(x_j)\in(0,1)$, where $F_j$ is the empirical distribution function of $x_j$ using mid-ranks to accommodate ties, and define $z_j=2ar_j-a$, with default $a=3$. This transformation requires no distributional assumption, reduces sparse tail cells, and places marginal occupancy on a comparable scale. Estimation and interpretation remain on the original predictor scale through the inverse mapping.

In Step~3, ROD selects lattice resolution \(L^*\) using separate fitting, validation, and test subsets, with allocations specified for each analysis. For candidate $L\in\mathcal{L}$, the grid contains $L^k$ cells and retains either the full contrast basis or contrasts through order $r$, with $p_L=\sum_{j=1}^{r}\binom{k}{j}(L-1)^j$. Candidates are completed before reconstruction, and only rank-admissible retained contrast systems are eligible for validation-based selection. The admissible candidate with the best validation performance is selected, while the test subset is reserved for final reporting.

In Steps~4--5, each probability-scaled predictor is partitioned into $L^*$ equal-population bins represented by equally spaced points on $z_j=2ar_j-a$, and observations are assigned to the nearest grid point in every dimension. If \(s_i\) is the observation's base weight and \(n^*=n/L^{*k}\) is the target total weight per cell, observation \(i\) in cell \(c(i)\) receives reconstruction weight \(w_i=n^*s_i/\sum_{h\in c(i)}s_h\), reducing in the unweighted case to \(w_i=n^*/n_{c(i)}\). Equalization across lattice cells defines the balanced product reference measure under which native contrasts and SS allocations are interpreted. Occupied cells retain their weighted observed means, while empty cells are passed to Step~6 for completion.

Step~6 completes the lattice without altering occupied-cell means. Let \(\mathcal S\) denote the occupied cells, with weighted means \(\mathbf y_{\mathcal S}\) and weights \(\mathbf W_{\mathcal S}\). Each empty cell \(g\notin\mathcal S\) receives \(\widetilde y_g=\mathcal C(g;\mathcal S)\), where \(\mathcal C\) is either a global main-effects WLS surface or a local neighbor-based completion rule. In the simulation, \((L,\mathcal C)\) is selected jointly by cross-validation.

For the selected \(L^*\), Step~7 constructs the orthogonal reconstruction and evaluates it through Freudenthal--Kuhn piecewise-affine interpolation \citep{Freudenthal1942,Kuhn1960}. Let \(\widetilde{\mathbf y}_L\) denote the observed and completed cell means, \(\mathbf X_L\) the retained contrast matrix, and \(\mathbf W_L\) the reconstruction weights. The contrast coefficients are \(\widehat{\boldsymbol\beta}_L=c_L^{-1}\mathbf X_L^{\top}\mathbf W_L\widetilde{\mathbf y}_L\), where \(\mathbf X_L^{\top}\mathbf W_L\mathbf X_L=c_L\mathbf I\). Within simplex \(s\), with vertices \(v_0,\ldots,v_k\), the surface is \(\widehat m(\mathbf x)=\sum_{r=0}^{k}\lambda_r(\mathbf x)\widehat y_{v_r}\), where \(\lambda_r(\mathbf x)\geq0\) and \(\sum_{r=0}^{k}\lambda_r(\mathbf x)=1\). The interpolant is continuous, parameter-free, exact at lattice vertices, and evaluable in \(O(k\log k)\) time \citep{Davies1996}; it does not alter the contrast coefficients, realized orthogonality, or SS decomposition.

Step~8 projects the reconstructed surface onto a declared structural basis, when one is known, and corrects finite-resolution recovery loss. Let \(\boldsymbol{\phi}(X)\) denote the target basis and \(\widehat{\boldsymbol\beta}_{\mathrm{raw}}\) its raw projection. The response-free calibration matrix \(\widehat{\mathbf C}_{L,k,X}\) is constructed by passing each pseudo-response \(y_i^{(j)}=\phi_j(X_i)\) through the fixed selected ROD pipeline while holding the mappings, lattice, completion rule, interpolation scheme, and projection points unchanged. When \(\widehat{\mathbf C}_{L,k,X}\) is full rank, \(\widehat{\boldsymbol\beta}=\widehat{\mathbf C}_{L,k,X}^{-1}\widehat{\boldsymbol\beta}_{\mathrm{raw}}\), with projected predictions \(\widehat m_{\phi}(X)=\boldsymbol{\phi}(X)^{\top}\widehat{\boldsymbol\beta}\). Calibration depends only on predictor geometry, lattice construction, interpolation, and the declared basis, not on the observed response.

\begin{figure}[H]
\centering{
\makebox[\textwidth][c]{\includegraphics[width=1.12\textwidth]{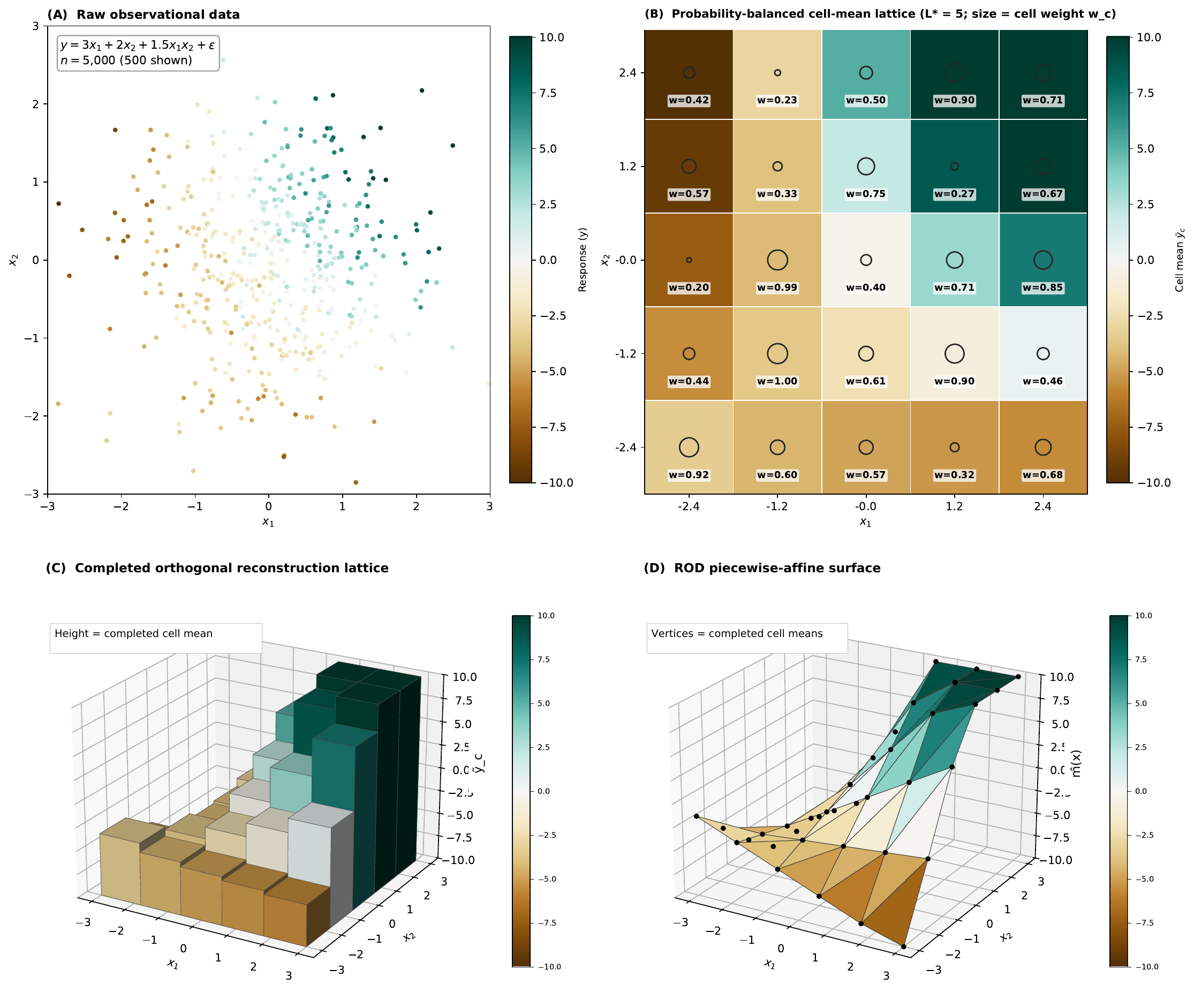}}
}
\caption{\label{fig-rod-emergence}Visualization of ROD for a low-dimensional interaction surface.}
\end{figure}

The same reconstructed surface may be projected onto alternative declared bases, each with its own response-free calibration operator, providing a sensitivity analysis of competing structural representations and effect hierarchies. When many effects or bases are tested, inference may proceed hierarchically \citep{Wu2009} with false-discovery control within stages \citep{Benjamini1995}.

\section{Theoretical Properties}\label{sec-properties}

The theoretical properties of ROD arise from the interaction between a balanced reference lattice and a mapped observational reconstruction. The framework yields algebraic orthogonality, specification-invariant attribution (within a fixed reconstruction's retained contrast space), order-independent SS, and a geometric representation of the reconstructed response surface.

\begin{theorem}[Reference-Lattice Orthogonality]
\label{theorem-orthogonality}
Let $\mathbf{X}\in\mathbb{R}^{L^k\times(L^k-1)}$ be the full nonintercept tensor-product ROD contrast matrix on a full, balanced $L^k$ lattice, with all cells retained at common total weight $n^*>0$. If the one-dimensional nonconstant contrasts are orthonormal and each interaction-order-$d_j$ column is scaled by $L^{(d_j-1)/2}$, then $\mathbf{W}=n^*\mathbf{I}_{L^k}$ and $\mathbf{X}^{\top}\mathbf{W}\mathbf{X}=n^*L^{k-1}\mathbf{I}=c\mathbf{I}$.
\textbf{Proof:} A full, balanced lattice gives $\mathbf W=n^*\mathbf I$. Tensor products of orthonormal one-dimensional contrasts are mutually orthogonal, so $\mathbf X^{\top}\mathbf X$ is diagonal. An order-$d_j$ column has unscaled squared norm $L^{k-d_j}$; scaling by $L^{(d_j-1)/2}$ makes every diagonal entry equal to $L^{k-1}$. Hence $\mathbf X^{\top}\mathbf W\mathbf X=n^*L^{k-1}\mathbf I=c\mathbf I$. $\blacksquare$
\end{theorem}

Theorem~\ref{theorem-orthogonality} gives the reference identity underlying ROD. The weighted Gram matrix has condition number one, and interaction-order normalization prevents higher-order contrast columns from receiving a different scale solely because they contain more nonconstant tensor factors.

\begin{corollary}[Specification and SS Invariance]
\label{cor-reference-projection}
Under Theorem~\ref{theorem-orthogonality}, with the weighted intercept fitted separately, any retained contrast block $\mathbf X_J$ has $\widehat{\boldsymbol{\beta}}_J=c^{-1}\mathbf X_J^{\top}\mathbf W\mathbf y$, independent of the other retained contrasts. Its Type~I, Type~II, and Type~III SS coincide and equal $SS_J=c^{-1}\lVert\mathbf X_J^{\top}\mathbf W\mathbf y\rVert_2^2$. \textbf{Proof:} Every retained subset inherits $\mathbf X_S^{\top}\mathbf W\mathbf X_S=c\mathbf I$. Because the contribution of $\mathbf X_J$ is unaffected by entry order or adjustment for other orthogonal blocks, its Type~I, Type~II, and Type~III SS coincide. $\blacksquare$
\end{corollary}

Theorem~\ref{theorem-orthogonality} establishes orthogonality for the full tensor-product basis. Any retained subset $\mathbf X_J\subseteq\mathbf X$ inherits $\mathbf X_J^{\top}\mathbf W\mathbf X_J=c\mathbf I$, so exact orthogonality, specification-invariant coefficients, and order-independent SS hold within any chosen contrast space.

\begin{proposition}[Conditional Lattice Error Representation]
\label{prop-cell-error}
Let the completed admissible lattice satisfy $\mathbf X^{\top}\mathbf W\mathbf X=c\mathbf I$, and write $\widetilde{\mathbf y}=\widetilde{\boldsymbol\mu}+\widetilde{\boldsymbol\varepsilon}$. If $\boldsymbol\beta^{*}=c^{-1}\mathbf X^{\top}\mathbf W\widetilde{\boldsymbol\mu}$, then $\widehat{\boldsymbol\beta}-\boldsymbol\beta^{*}=c^{-1}\mathbf X^{\top}\mathbf W\widetilde{\boldsymbol\varepsilon}$. Thus, conditional on the selected lattice, completion rule, provenance, and weights, coefficient error is determined by the reconstructed lattice errors. \textbf{Proof:} Substitute $\widetilde{\mathbf y}=\widetilde{\boldsymbol\mu}+\widetilde{\boldsymbol\varepsilon}$ into $\widehat{\boldsymbol\beta}=c^{-1}\mathbf X^{\top}\mathbf W\widetilde{\mathbf y}$. $\blacksquare$
\end{proposition}

\begin{proposition}[Pure-Error Invariance and Unbiasedness]
\label{prop-pure-error}
For fixed cell assignments, let $SS_E=\sum_{c:n_c>1}\sum_{i=1}^{n_c}(y_{ic}-\bar y_c)^2$, $\nu_E=\sum_{c:n_c>1}(n_c-1)$, and $MS_E=SS_E/\nu_E$, with $\nu_E>0$. Then $MS_E$ is invariant to the retained contrast specification. If $y_{ic}=\mu_c+\varepsilon_{ic}$ with independent mean-zero errors of variance $\sigma^2$, then $E(MS_E)=\sigma^2$. This result applies only to observed replicated cells and does not quantify uncertainty from completion, calibration, or lattice selection. \textbf{Proof:} $SS_E$ depends only on within-cell deviations. Under the stated error model, $E(SS_E)=\nu_E\sigma^2$, so $E(MS_E)=\sigma^2$. Singleton cells contribute no within-cell degrees of freedom. $\blacksquare$
\end{proposition}

Fixing the admissible lattice and completed cell means gives ROD a dual identity: an orthogonal contrast system and a reconstructed response surface. When a structural basis is declared, calibrated projection provides an additional finite-dimensional representation of that fixed surface.

\begin{proposition}[Exact recovery and calibration stability]
\label{prop-calibration}
Fix the reconstruction pipeline and let
$\mathcal{R}:\mathbb{R}^{m}\rightarrow\mathbb{R}^{p}$ denote the resulting
linear raw-recovery operator evaluated at $m$ projection points. Let
$\boldsymbol{\Phi}\in\mathbb{R}^{m\times p}$ contain the declared basis
functions evaluated at those points, and define the calibration matrix
$\mathbf{C}=\mathcal{R}\boldsymbol{\Phi}$. If $\mathbf{C}$ is nonsingular,
the calibrated projection is
$\widehat{\boldsymbol{\beta}}
=\mathbf{C}^{-1}\mathcal{R}\mathbf{y}$.
If the noiseless response lies in the declared basis,
$\mathbf{y}=\boldsymbol{\Phi}\boldsymbol{\beta}$, then
$\widehat{\boldsymbol{\beta}}=\boldsymbol{\beta}$. More generally, if
$\mathbf{y}=\boldsymbol{\Phi}\boldsymbol{\beta}+\boldsymbol{\varepsilon}$,
then
$\widehat{\boldsymbol{\beta}}-\boldsymbol{\beta}
=\mathbf{C}^{-1}\mathcal{R}\boldsymbol{\varepsilon}$ and
$\|\widehat{\boldsymbol{\beta}}-\boldsymbol{\beta}\|
\leq
\|\mathbf{C}^{-1}\|
\|\mathcal{R}\boldsymbol{\varepsilon}\|$. \textbf{Proof:} Because $\mathbf{C}=\mathcal{R}\boldsymbol{\Phi}$,
$\widehat{\boldsymbol{\beta}}
=\mathbf{C}^{-1}\mathcal{R}\boldsymbol{\Phi}\boldsymbol{\beta}
=\boldsymbol{\beta}$ under the declared basis. Substitution of
$\mathbf{y}=\boldsymbol{\Phi}\boldsymbol{\beta}
+\boldsymbol{\varepsilon}$ gives the error identity, and the norm bound
follows from submultiplicativity.
\end{proposition}

Proposition~\ref{prop-calibration} shows that response-free calibration removes the deterministic recovery distortion induced by a fixed reconstruction pipeline when the surface lies in the declared basis. Calibration does not eliminate reconstruction, completion, approximation, or sampling error, and poor conditioning of $\mathbf{C}$ can amplify the projection error. Structural coefficients and marginal effects are derived summaries of the reconstructed surface, whereas SS arise from its native orthogonal contrasts. Projection does not alter the reconstructed lattice, cell provenance, orthogonality, or SS decomposition.

\section{Simulation}\label{sec-simulation-results}

A simulation assessed whether finite-sample ROD preserves reference-grid orthogonality, predictive competitiveness, and original-scale coefficient recovery under a deployed pipeline selector. Each condition used a 60\%/20\%/20\% train-validation-test split. Resolution and completion, \((L^*,\mathcal{C}^*)\), were selected by validation performance among rank-admissible candidates using mappings, cell assignments, and completion estimated from the training subset only. The selected pipeline was then refit on the combined training and validation subsets and evaluated once on an untouched test set; neither test predictors nor test responses entered pipeline fitting or selection.

The simulation crossed \(k\in\{3,4,5,6\}\), \(n\in\{0.5,\,2.5,\,10,\,50,\,150,\,500\}\) thousand observations, \(\rho\in\{0.0,0.4,0.8\}\), nine data-generating processes (DGPs; Table~\ref{tbl-dgp}), and \(R=10\) replications, producing \(6{,}480\) conditions. Candidate resolutions \(L\in\{3,\ldots,8\}\) were evaluated within each condition using two prespecified completion rules: a global main-effects surface and local inverse-distance weighting. ROD retained all one-dimensional contrasts and all two-factor tensor-product contrasts, corresponding to interaction order \(r=2\), throughout the simulation.

Within each condition, observations were generated independently with \(\mathbf{x}_i\sim N_k(\mathbf{0},\boldsymbol{\Sigma}_{\rho})\). The covariance matrix had unit variances and off-diagonal correlations \(\rho,\rho(0.5),\rho(0.5)^2,\ldots\), assigned to predictor pairs in the order \((1,2),(1,3),\ldots,(k-1,k)\) and verified to be positive semidefinite across the full simulation grid. Unless stated otherwise, \(\varepsilon_i\sim N(0,1)\) independently of \(\mathbf{x}_i\). For the hurdle DGP, \(u_i\sim\operatorname{Uniform}(0,1)\) and \(\eta_i,\varepsilon_i\sim N(0,1)\) were mutually independent and independent of \(\mathbf{x}_i\). Because \(E[\exp(0.30\eta_i-0.045)]=1\), \(E(y_i\mid\mathbf{x}_i)=p_{+,i}\mu_{+,i}\). The simulation was implemented in Python \citep{Python}.

\begin{table}[H]
\centering
\caption{Data-generating processes used in the main simulation study.}
\label{tbl-dgp}
\small
\begin{tabular}{ll}
\toprule
\textbf{Process} & \textbf{Response model} \\
\midrule
Linear (LINEAR) & \(y=3x_1+2x_2+\varepsilon\) \\
Quadratic + Interaction (QUAD) & \(y=3x_1+2x_2+1.5\,x_1x_2+1.0\,x_1^2+\varepsilon\) \\
Log saturation (LOGSAT) & \(y=3\,\mathrm{softplus}(x_1)+2\,\mathrm{softplus}(x_2)+0.5x_3+\varepsilon\) \\
Exponential growth (EXP) & \(y=2\exp(0.35x_1)+1.5\exp(0.25x_2)+0.5x_3+\varepsilon\) \\
Piecewise linear (PIECEWISE) & \(y=1.5x_1+2.5\max(x_1-0.5,0)-1.0\max(-x_2-0.25,0)+0.5x_3+\varepsilon\) \\
Strong threshold (THRESHOLD) & \(y=3.0\,\mathbf{1}(x_1{>}0)+2.5\,\mathbf{1}(x_2{>}0)+2.0\,\mathbf{1}(x_1{>}0)\mathbf{1}(x_2{>}0)+\varepsilon\) \\
Two-part hurdle (HURDLE) & \(p_+=\mathrm{logit}^{-1}(x_1-0.75x_2+0.5x_3),\quad d=\mathbf{1}(u<p_+)\) \\
& \(y=d\,\mu_+\exp(0.30\eta-0.045)+0.25\varepsilon,\quad \mu_+=\exp(0.50+0.55x_1+0.25x_2)\) \\
XOR interaction (XOR)& \(y=4.0\cdot\mathbf{1}\{\mathrm{sign}(x_1)=\mathrm{sign}(x_2)\}-2.0+0.5x_3+\varepsilon\) \\
Bump (BUMP) & \(y=5.0\exp\!\left(-\tfrac{(x_1-1)^2+(x_2+1)^2}{0.5}\right)+0.5x_3+\varepsilon\) \\
\bottomrule
\end{tabular}

\begin{flushleft}
\scriptsize \(\operatorname{softplus}(t)=\log(1+\exp(t))\), \(\operatorname{logit}^{-1}(t)=[1+\exp(-t)]^{-1}\), and \(\mathbf{1}(\cdot)\) denotes the indicator function.
\end{flushleft}
\end{table}

\subsection{Reconstruction Admissibility, Completion, and Resolution}\label{sec-sim-primary}

Across all \(6{,}480\) planned conditions, every condition produced an estimable reconstruction, with no computational errors or unintended skips. Resolution selection was constrained to $L^*=\{3 \ldots 8\}$ to control computational complexity. The most frequently selected resolution was \(L^*=8\), which occurred in \(2{,}838\) conditions (43.8\%), followed by \(L^*=4\) in \(1{,}490\) conditions (23.0\%), \(L^*=6\) in \(653\) conditions (10.1\%), \(L^*=7\) in \(648\) conditions (10.0\%), \(L^*=5\) in \(456\) conditions (7.0\%), and \(L^*=3\) in \(395\) conditions (6.1\%).

Validation selected global main-effects completion in \(4{,}624\) conditions (\(71.4\%\)) and local inverse-distance weighting (IDW), which fills an empty cell using a distance-weighted average of nearby occupied-cell means, in \(1{,}856\) conditions (\(28.6\%\)). Exact validation ties occurred in \(2{,}211\) conditions (\(34.1\%\)) and were resolved by the deterministic selection rule. IDW selection ranged from \(3.3\%\) for LINEAR to \(70.0\%\) for QUAD.

\subsection{Prediction Performance}\label{sec-sim-prediction}

For predictive comparisons, ROD was evaluated on the same declared structural basis as the parametric comparator when prespecified. Accordingly, the LINEAR and QUAD DGPs use calibrated ROD projections onto the OLS-ME and POLY bases, respectively, while the remaining DGPs use predictions from ROD's surface. Table~\ref{tbl-r2-by-dgp-k} reports mean out-of-sample $R^2$ by DGP. OLS-ME is included as a naive baseline, quantifying the cost of ignoring nonlinear or interaction structure.

For the LINEAR DGP, calibrated ROD and OLS-ME produced nearly identical mean performance. ROD exceeded POLY by a mean \(\Delta R^2=0.0005\) (stratified paired bootstrap 95\% CI: \(0.0004\) to \(0.0006\)) and achieved higher \(R^2\) in \(69.7\%\) of matched conditions (exact paired sign test, Holm-adjusted \(p<.001\)). For QUAD, calibrated ROD and POLY were practically indistinguishable, with a mean difference of only \(\Delta R^2=0.0001\), although ROD achieved higher \(R^2\) in \(55.7\%\) of matched conditions (Holm-adjusted exact sign-test \(p=.003\)). ROD substantially outperformed POLY for THRESHOLD, XOR, and BUMP, with mean \(R^2\) gains of \(0.2201\), \(0.3163\), and \(0.0933\), respectively, and the proportion favoring ROD exceeding \(92\%\). POLY performed better for PIECEWISE, LOGSAT, EXP, and HURDLE, for which ROD's mean \(R^2\) differences ranged from \(-0.0399\) to \(-0.0870\).

\begin{table}[H]
\centering
\caption{Mean out-of-sample \(R^2\) and paired ROD--POLY differences by DGP. Difference intervals use a stratified paired bootstrap over \((n,k,\rho)\) cells; sign-test \(p\)-values are Holm--Bonferroni adjusted.}
\label{tbl-r2-by-dgp-k}
\scriptsize
\resizebox{\textwidth}{!}{%
\begin{tabular}{lccccc}
\toprule
DGP & OLS-ME & POLY & ROD & Mean $\Delta R^2$ (95\% CI) & \% \(\Delta R^2>0\) (\(p\), Holm) \\
\midrule
QUAD
& 0.6831 \((0.6800,\ 0.6862)\)
& 0.9579 \((0.9575,\ 0.9583)\)
& 0.9580$^*$ \((0.9576,\ 0.9584)\)
& $+0.0001$ \((>0.0000,\ 0.0002)\)
& 55.7\% (0.003)
\\
LINEAR
& 0.9438 \((0.9434,\ 0.9441)\)
& 0.9433 \((0.9429,\ 0.9437)\)
& 0.9438$^*$ \((0.9434,\ 0.9441)\)
& $+0.0005$ \((+0.0004,\ +0.0006)\)
& 69.7\% ($<0.001$)
\\
THRESHOLD
& 0.6009 \((0.5992,\ 0.6025)\)
& 0.6051 \((0.6032,\ 0.6070)\)
& 0.8252 \((0.8239,\ 0.8265)\)
& $+0.2201$ \((+0.2185,\ +0.2217)\)
& 100.0\% ($<0.001$)
\\
PIECEWISE
& 0.8172 \((0.8161,\ 0.8184)\)
& 0.8631 \((0.8620,\ 0.8641)\)
& 0.8200 \((0.8185,\ 0.8215)\)
& $-0.0430$ \((-0.0444,\ -0.0417)\)
& 1.1\% ($<0.001$)
\\
LOGSAT
& 0.7802 \((0.7789,\ 0.7815)\)
& 0.8336 \((0.8323,\ 0.8347)\)
& 0.7893 \((0.7880,\ 0.7907)\)
& $-0.0442$ \((-0.0454,\ -0.0432)\)
& 0.3\% ($<0.001$)
\\
XOR
& 0.0468 \((0.0450,\ 0.0486)\)
& 0.2868 \((0.2836,\ 0.2898)\)
& 0.6030 \((0.6008,\ 0.6052)\)
& $+0.3163$ \((+0.3132,\ +0.3195)\)
& 99.4\% ($<0.001$)
\\
EXP
& 0.5488 \((0.5465,\ 0.5510)\)
& 0.5648 \((0.5624,\ 0.5671)\)
& 0.5249 \((0.5224,\ 0.5273)\)
& $-0.0399$ \((-0.0411,\ -0.0386)\)
& 2.8\% ($<0.001$)
\\
HURDLE
& 0.3422 \((0.3393,\ 0.3450)\)
& 0.4517 \((0.4467,\ 0.4565)\)
& 0.3647 \((0.3614,\ 0.3680)\)
& $-0.0870$ \((-0.0908,\ -0.0832)\)
& 5.1\% ($<0.001$)
\\
BUMP
& 0.2396 \((0.2370,\ 0.2422)\)
& 0.2682 \((0.2648,\ 0.2716)\)
& 0.3615 \((0.3586,\ 0.3644)\)
& $+0.0933$ \((+0.0911,\ +0.0956)\)
& 92.2\% ($<0.001$)
\\
\bottomrule
\end{tabular}
}
\begin{flushleft}
\footnotesize * ROD calibrated to the prespecified basis. Mean \(\Delta R^2\): stratified paired bootstrap.
\end{flushleft}
\end{table}

As a robustness check, mean $R^2$ was also examined disaggregated by $k \in \{3,4,5,6\}$ and $\rho \in \{0, 0.4, 0.8\}$. Neither dimension produced a substantively meaningful effect: across all DGP--model combinations, the largest observed half-range in mean $R^2$ across $k$ was $0.047$ (ROD, XOR interaction), and the largest across $\rho$ was $0.081$ (ROD, XOR interaction), against an $R^2$ scale of $[0,1]$. The proportion of individual conditions in which ROD outperformed POLY was similarly stable across $\rho$ for most DGPs, with two exceptions: for XOR interaction and Bump, ROD's $R^2$ performance declined moderately as $\rho$ increased (XOR interaction: $100.0\%$, $99.2\%$, $99.2\%$ at $\rho = 0, 0.4, 0.8$; Bump: $98.8\%$, $96.7\%$, $81.2\%$), while remaining decisively above chance throughout.

\subsection{Recovery of Prespecified Structural Coefficients}\label{sec-sim-coef}

Coefficient recovery was evaluated for four DGPs with prespecified structural targets. Linear and Quadratic+Interaction used continuous polynomial bases, while Strong Threshold and XOR used positive-indicator and sign-interaction bases. These bases were introduced only after predictive fitting, and ROD coefficients were recovered using the Step~8 calibration operator.

Table~\ref{tbl-coef-bias} shows that calibrated ROD bias remained near zero and dispersion comparatively stable across targets and dimensions. OLS and POLY are reported only when their specifications contain the corresponding structural term; the Strong Threshold and XOR results therefore assess conditional recovery under a prespecified nonlinear basis, not functional-form discovery. For Quadratic+Interaction \(X_1X_2\), POLY bias changed from \(+0.002\) at \(k=3\) to \(-0.031\) at \(k=6\), while its SD increased from \(0.046\) to \(0.124\); ROD bias remained near zero and its SD increased only from \(0.042\) to \(0.067\).

\begin{table}[H]
\centering
\caption{ROD coefficient bias and standard deviation by \(k\) with parametric benchmarks where available.}
\label{tbl-coef-bias}
\scriptsize
\begin{tabular}{lrrrr}
\toprule
& \multicolumn{4}{c}{\(k\)} \\
\cmidrule(lr){2-5}
Target (method) & 3 & 4 & 5 & 6 \\
\midrule
Linear \(X_1\) (ROD)
& +0.004 (0.031) & \(-0.001\) (0.034) & +0.003 (0.036) & +0.002 (0.030) \\
Linear \(X_1\) (OLS)
& +0.001 (0.025) & \(-0.001\) (0.028) & +0.003 (0.034) & +0.003 (0.032) \\
\midrule
Quadratic+Interaction \(X_1X_2\) (ROD)
& \(-0.002\) (0.042) & \(-0.003\) (0.043) & \(-0.001\) (0.042) & +0.001 (0.067) \\
Quadratic+Interaction \(X_1X_2\) (POLY)
& +0.002 (0.046) & +0.013 (0.079) & \(-0.027\) (0.122) & \(-0.031\) (0.124) \\
\midrule
Strong Threshold \(X_1X_2\) (ROD)
& +0.014 (0.132) & +0.002 (0.127) & +0.003 (0.106) & \(-0.022\) (0.138) \\
\midrule
XOR \(X_1X_2\) (ROD)
& \(-0.002\) (0.028) & \(-0.001\) (0.032) & +0.001 (0.032) & 0.000 (0.026) \\
\bottomrule
\end{tabular}
\begin{flushleft}
\footnotesize OLS and POLY are reported only where their declared specifications contain the same structural coefficient as ROD.
\end{flushleft}
\end{table}

\subsection{Analyst Operating Envelope and Sample-Size Guidance}\label{sec-operator-envelope}

Under independence, the structural ROD sample-size benchmark is Rao's lower bound for an equal-level strength-four orthogonal array \citep{Rao1947}, \(N_R=1+k(L-1)+\binom{k}{2}(L-1)^2\). Let \(M_{\Sigma}=E_{\Sigma}\{\phi(X)\phi(X)^{\top}\}\) denote the information matrix for the retained contrast basis and let \(M_0\) denote the corresponding information matrix under independence. The dependence-adjusted sample size is defined by requiring the dependent design to provide at least the Rao benchmark information in every retained direction, \(nM_{\Sigma}\succeq N_R M_0\), where \(\succeq\) denotes the Loewner ordering \citep{Pukelsheim2006}. Defining the relative information matrix \(R_{\Sigma}=M_0^{-1/2}M_{\Sigma}M_0^{-1/2}\), this condition is equivalent to \(n\lambda_{\min}(R_{\Sigma})\ge N_R\), yielding \(N_{R,\Sigma}=\left\lceil N_R/\lambda_{\min}(R_{\Sigma})\right\rceil\). Under independence and orthonormal contrast coding, \(M_0=I\), so the adjustment reduces to \(N_R\). For the reduced case in which \(\phi(X)\) contains standardized linear main effects only, \(M_{\Sigma}=\Sigma\), giving \(N_{R,\Sigma}=\left\lceil N_R/\lambda_{\min}(\Sigma)\right\rceil\). The minimum relative eigenvalue is also the E-optimality criterion and represents the least-supported retained information direction. This construction is an information-matching adjustment to Rao's combinatorial lower bound for planning purposes, not a new Rao bound or an orthogonal-array existence result. See Table~\ref{tbl-rao-corr} for sample-size considerations.

\begin{table}[H]
\centering
\caption{Rao-based ROD sample-size benchmarks for structural order \(r=2\). Resolution V values are shown first with Resolution IV in parentheses; dependence adjustment uses \(1/(1-\bar{\rho})\).}
\label{tbl-rao-corr}
\begin{scriptsize}
\begin{tabular}{rrr|rrr}
\hline
\(k\) & \(L\) & \(N_R\) &
\(\bar{\rho}=.25\) & \(\bar{\rho}=.50\) & \(\bar{\rho}=.75\) \\
& & \(1.00\times\) &
\(1.33\times\) & \(2.00\times\) & \(4.00\times\) \\
\hline
5  & 5  & 181 (85)     & 242 (114)    & 362 (170)    & 724 (340) \\
5  & 10 & 856 (370)    & 1142 (494)   & 1712 (740)   & 3424 (1480) \\
5  & 15 & 2031 (855)   & 2708 (1140)  & 4062 (1710)  & 8124 (3420) \\
5  & 20 & 3706 (1540)  & 4942 (2054)  & 7412 (3080)  & 14824 (6160) \\
\hline
10 & 5  & 761 (185)    & 1015 (247)   & 1522 (370)   & 3044 (740) \\
10 & 10 & 3736 (820)   & 4982 (1094)  & 7472 (1640)  & 14944 (3280) \\
10 & 15 & 8961 (1905)  & 11948 (2540) & 17922 (3810) & 35844 (7620) \\
10 & 20 & 16436 (3440) & 21915 (4587) & 32872 (6880) & 65744 (13760) \\
\hline
15 & 5  & 1741 (285)   & 2322 (380)   & 3482 (570)   & 6964 (1140) \\
15 & 10 & 8641 (1270)  & 11522 (1694) & 17282 (2540) & 34564 (5080) \\
15 & 15 & 20791 (2955) & 27722 (3940) & 41582 (5910) & 83164 (11820) \\
15 & 20 & 38191 (5340) & 50922 (7120) & 76382 (10680)& 152764 (21360) \\
\hline
20 & 5  & 3121 (385)   & 4162 (514)   & 6242 (770)   & 12484 (1540) \\
20 & 10 & 15571 (1720) & 20762 (2294) & 31142 (3440) & 62284 (6880) \\
20 & 15 & 37521 (4005) & 50028 (5340) & 75042 (8010) & 150084 (16020) \\
20 & 20 & 68971 (7240) & 91962 (9654) & 137942 (14480)& 275884 (28960) \\
\hline
\end{tabular}
\end{scriptsize}
\end{table}

\subsection{Software Implementation}\label{sec-software}

ROD is implemented as an operational Python software package that executes the reconstruction procedure described above. The implementation includes probability-balanced lattice construction, preservation of supported weighted cell means, completion of unsupported cells, tensor-product orthogonal contrast construction, validation-based joint selection of lattice resolution and completion rule, Freudenthal piecewise-affine evaluation, response-free projection calibration, weighted estimation, categorical adjustment and stratification, and numerical diagnostics for rank and conditioning. The implementation separates the user-facing ROD estimator from the simulation and empirical-analysis code used in this paper.

The version used to validate the implementation is ROD~0.4. The package source code, installation materials, tests, examples, and empirical golden-equivalence validation are maintained at \url{https://github.com/dustoff06/ROD}. Manuscript-specific simulation, bootstrap, and empirical-analysis code and the generated results are maintained in the same repository as replication materials. Thus, the software implementation of ROD and the code used to evaluate the method in this study are independently identifiable but reproducible from a versioned repository.

\section{Empirical Illustration: The Mincer Earnings Equation}\label{sec-mincer}

We illustrate ROD using the Mincer earnings equation \citep{Mincer1974}, \(\ln(\mathrm{earnings}_i)=\beta_0+\beta_1\mathrm{schooling}_i+\beta_2\mathrm{experience}_i+\beta_3\mathrm{experience}_i^2+\varepsilon_i\). Schooling and potential experience define the two-dimensional ROD grid. The sample is restricted to male, married, White respondents with positive annual wage and salary income to maintain a homogeneous illustrative subgroup.

The analysis uses the 2018 CPS Annual Social and Economic Supplement from IPUMS CPS \citep{IPUMSCPS2024}, with \(n=21{,}226\) observations after restriction. All probability mappings, estimation, model selection, and evaluation use the person weight \texttt{ASECWT}, normalized to mean one. The outcome is log annual wage and salary income, and potential experience equals age minus schooling minus six, truncated at zero. An 80\%/20\% split reserves \(4{,}245\) observations for testing. The remaining \(16{,}981\) observations are split 80\%/20\% for training and validation, with \(L^*\) chosen to maximize weighted validation \(R^2\) among rank-admissible candidates.

\begin{figure}[H]
\centering{
\makebox[\textwidth][c]{\includegraphics[width=.6\textwidth]{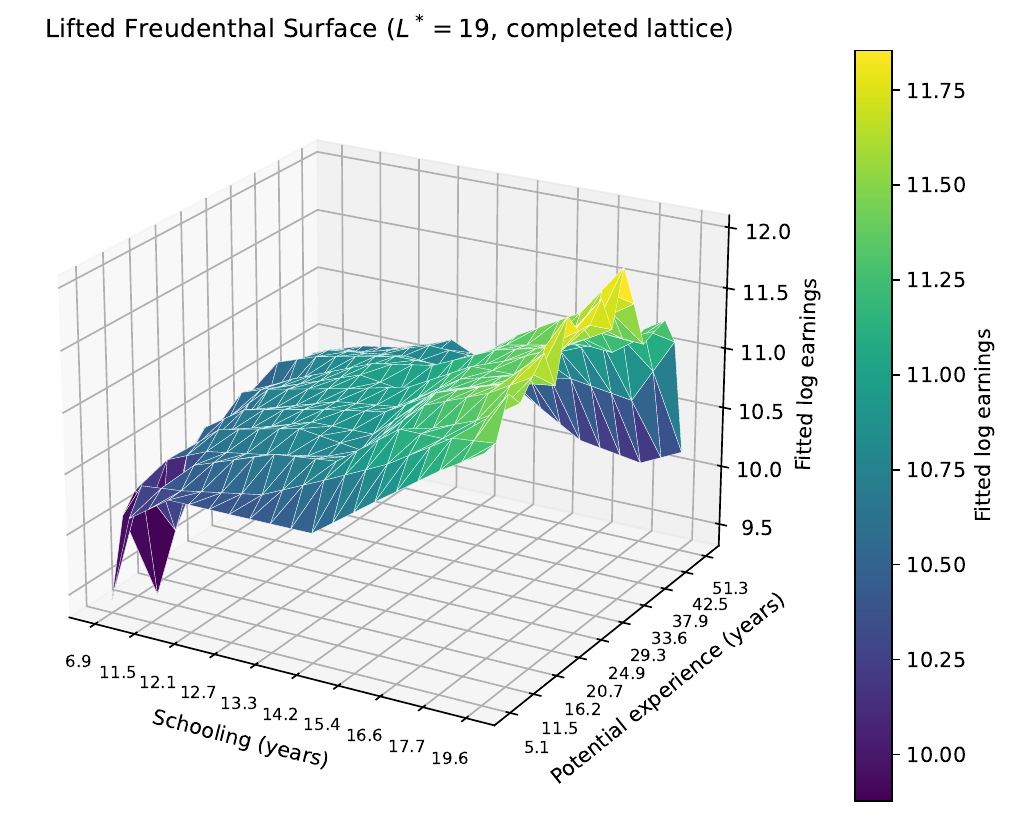}}
}
\caption{\label{fig-mincer}ROD reconstruction of the Mincer schooling--experience surface.}
\end{figure}

Under the Step~2 probability-scale mapping with mid-rank handling of ties, the 14 schooling levels occupied eight columns under the CV-selected \(L^*=19\), leaving 11 structurally unsupported schooling columns (\(209\) of \(361\) cells) in Figure~\ref{fig-mincer}. These cells were completed using the Step~6 local rule, specialized to linear interpolation along schooling between bracketing occupied cells, while occupied cells retained their \texttt{ASECWT}-weighted means. The same completion rule was used during CV and refit on the non-test sample. After completion and equalization of reconstruction weight across lattice cells, the reconstruction yielded \(\kappa=1.000\) and \(\max|\operatorname{offdiag}(\mathbf{X}^{\top}\mathbf{W}\mathbf{X})|=1.16\times10^{-6}\), confirming orthogonal construction.

Table~\ref{tbl-native-r2} reports weighted out-of-sample \(R^2\) with conditional 95\% bootstrap confidence intervals (CI), computed from \(5{,}000\) paired resamples of the \(4{,}245\) held-out test observations while retaining their survey weights. OLS-ME fits schooling and experience without curvature or interaction terms. OLS-Mincer adds the conventional quadratic experience term. POLY fits the full second-order polynomial, adding schooling squared and the schooling-by-experience interaction. ROD reports predictions from the native piecewise-affine reconstruction without requiring a declared structural basis.

ROD produced the highest weighted out-of-sample \(R^2\) point estimate, followed by POLY. Their estimates differed by only \(0.0025\), and their confidence intervals overlapped substantially. Both outperformed the conventional OLS-Mincer specification in point estimates, while OLS-ME provided the expected lower-dimensional baseline.

\begin{table}[H]
\centering
\caption{Weighted out-of-sample predictive performance.}
\label{tbl-native-r2}
\small
\begin{tabular}{lcc}
\toprule
Method & \(R^2_{\text{test}}\) & 95\% CI \\
\midrule
OLS-ME (schooling + experience)
& 0.1141 & \((0.0913,\ 0.1375)\) \\
OLS-Mincer (schooling + experience + experience squared)
& 0.1586 & \((0.1345,\ 0.1840)\) \\
POLY (linear, squared, and interaction terms)
& 0.1799 & \((0.1555,\ 0.2053)\) \\
ROD (native reconstructed surface)
& 0.1824 & \((0.1548,\ 0.2104)\) \\
\bottomrule
\end{tabular}
\end{table}

Table~\ref{tbl-mincer-coefficients} compares directly estimated OLS coefficients with POLY and calibrated ROD projections onto the Mincer basis. The POLY and ROD projections use the same equally weighted Freudenthal facet centroids. All three imply positive schooling effects and concave experience profiles.

\begin{table}[H]
\centering
\caption{Mincer-basis coefficients for OLS, POLY, and ROD. NOTE: Peak experience computed from unrounded coefficients.}
\label{tbl-mincer-coefficients}
\small
\begin{tabular}{lrrrr}
\toprule
Method & Schooling & Exp. & Exp.$^2$ & Peak Exp. \\
\midrule
OLS (linear + squared term) & 0.0974 & 0.0576 & $-0.00109$ & 26.51 \\
POLY (linear + squared + interaction)    & 0.1083 & 0.0661 & $-0.00125$ & 26.43 \\
ROD (Mincer basis declared)               & 0.1136 & 0.0620 & $-0.00117$ & 26.56 \\
\bottomrule
\end{tabular}
\end{table}

Table~\ref{tbl-mincer-decomp-side-by-side} compares ROD's order-invariant unique SS shares with the ranges of sequential SS allocations obtained across all \(3!=6\) OLS orderings and \(5!=120\) POLY orderings. ROD orthogonality makes the five effect-family allocations non-overlapping and exhaustive, whereas the OLS and POLY allocations vary with term-entry order.

\begin{table}[H]
\centering
\caption{Order-invariant ROD unique SS shares and order-dependent sequential SS ranges for the weighted CPS Mincer example, \(L^*=19\).}
\label{tbl-mincer-decomp-side-by-side}
\small
\begin{tabular}{lrrr}
\toprule
Effect family & ROD unique share & OLS sequential range & POLY sequential range \\
\midrule
Schooling (linear)
& 58.19\%
& 61.04--70.39\%
& 0.00--64.28\% \\
Schooling (nonlinear)
& 2.86\%
& ---
& 0.46--65.83\% \\
Experience (linear)
& 0.71\%
& 1.06--24.21\%
& 0.09--45.65\% \\
Experience (nonlinear)
& 27.40\%
& 5.39--32.54\%
& 4.91--64.27\% \\
Schooling \(\times\) experience
& 10.84\%
& ---
& 0.38--52.22\% \\
\bottomrule
\end{tabular}
\begin{flushleft}
\footnotesize SS Shares describe within-model allocation only. ROD nonlinear includes all higher-order orthogonal contrasts; OLS and POLY entries represent squared terms.
\end{flushleft}
\end{table}

The illustration distinguishes the native ROD surface, its calibrated coefficient projection, and the OLS and POLY comparators. ROD does not replace the Mincer equation or resolve causal identification concerns such as ability bias \citep{Card1999}. Its contribution is an interpretable reconstruction with orthogonal attribution, explicit structural diagnostics, competitive out-of-sample performance, and substantively plausible projected coefficients.

\section{Discussion}\label{sec-discussion}

ROD's main contribution is stable attribution under correlated observational predictors. Once an admissible reference lattice and retained contrast space are fixed, \(\mathbf{X}^{\top}\mathbf{W}\mathbf{X}=c\mathbf{I}\) yields specification-invariant contrasts and unique, order-independent SS under the declared reference measure \citep{montgomery2019design}. Attribution therefore becomes a reproducible property of the reconstruction rather than a consequence of term order or the inclusion of other retained orthogonal effects. Interpolation affects only between-lattice evaluation, while calibration expresses the reconstructed surface in a declared scientific basis without altering its orthogonality or SS decomposition.

The simulations show that this stability does not require sacrificing predictive performance uniformly, but neither does ROD dominate conventional regression. ROD was essentially equivalent to POLY for linear and quadratic-interaction surfaces, performed substantially better for threshold, sign-interaction, and localized structures, and performed worse for several smoother or distributionally complex processes. These results suggest that ROD is most useful when flexible surface reconstruction and stable attribution are both important. Its advantage is not universal predictive superiority, but the ability to recover interpretable structure without making attribution depend on a conventional regression specification. Calibrated coefficient recovery further showed that the reconstructed surface can be mapped accurately to prespecified structural bases, although such recovery is conditional on the basis being declared rather than discovered automatically.

The CPS illustration makes this distinction concrete. ROD and POLY produced similar predictive performance and similar Mincer-basis projections, so the empirical result is not evidence of a decisive forecasting advantage. The difference lies in attribution: sequential OLS and POLY SS allocations varied substantially with term order despite identical fitted models, whereas ROD produced a single non-overlapping and exhaustive allocation under the declared reconstruction. ROD therefore changes the interpretive question from which specification or ordering receives credit to how variation is allocated within a defined reference surface.

That reference surface is central to interpretation. ROD defines attribution under a balanced product reference measure rather than the empirical predictor distribution, making the measure explicit and reproducible across samples and replications. Completion is part of the estimator rather than a computational repair, since incomplete support requires interpolation, extrapolation, pooling, or omission in any response-surface method. ROD makes these choices visible through preserved observed-cell means, provenance, completion diagnostics, and validation-based pipeline selection. Exact orthogonality does not imply that completed cells equal their unobserved population means, so support, completion burden, conditioning, and provenance remain essential diagnostics. ROD is therefore most compelling when interpretive stability, reproducibility, and auditable attribution matter alongside prediction, and less so when prediction alone is the objective or a low-dimensional parametric model is sufficient. It remains representational rather than causal and does not resolve identification problems such as ability bias in the Mincer application \citep{Card1999}.

\section{Conclusion}\label{sec-conc}

ROD reframes analysis of correlated observational predictors as a reconstruction problem rather than an attempt to recover orthogonality from the observed design matrix. By constructing an admissible balanced reference surface first, ROD makes attribution a property of a declared reconstruction rather than of term order or a particular regression specification. The resulting contrasts and SS therefore provide a common basis for comparing effects across specifications, samples, and analysts, while interpolation and calibrated projection allow the same reconstructed surface to support prediction and interpretation in scientifically meaningful coordinates.

The empirical and simulation results show that this interpretive stability need not come at the expense of useful predictive performance, but prediction is not ROD's primary advantage. Conventional parametric models remain effective when their functional form is appropriate, while ROD becomes more valuable when nonlinear structure, interactions, incomplete support, or correlated predictors make attribution sensitive to modeling choices. Its contribution is therefore not that it universally predicts better, but that it separates surface reconstruction from subsequent attribution and structural interpretation, making those steps explicit, reproducible, and auditable.

That transparency also defines the limits of the method. ROD's conclusions are conditional on the declared reference measure, available support, completion rule, retained contrast space, and structural basis used for projection. Orthogonality cannot create information where observations are absent, and completion remains an assumption about the unobserved surface rather than a substitute for data. Within those boundaries, ROD provides a practical framework for observational analyses in which prediction alone is insufficient and scientific conclusions require stable, comparable, and traceable explanations of variation.

\section{Disclosure statement}\label{disclosure-statement}

This research received no external funding. The authors declare no conflicts of interest.

\section{Data Availability Statement}\label{data-availability-statement}

The operational Python implementation of ROD and manuscript replication materials are available at \url{https://github.com/dustoff06/ROD}. Upon acceptance, the versioned software, analysis code, and frozen results will be archived there; public empirical datasets are cited in the manuscript with access instructions provided in the repository.

\noindent\textbf{Acknowledgments.} The authors acknowledge the use of large language model (LLM) systems, including ChatGPT 5.5 (OpenAI) and Claude Sonnet 5 (Anthropic), as interactive research and writing assistants during manuscript development.

\noindent\textbf{Author Contributions.} L.V.F., C.P.F., A.S., and A.T. contributed to conceptualization, methodology, validation, formal analysis, investigation, resources, and manuscript review and editing. All authors read and approved the manuscript.

\noindent\textbf{Disclaimer.} The views expressed are those of the authors and do not necessarily reflect those of their affiliated institutions. Generative artificial intelligence systems are not authors, and responsibility for the manuscript rests solely with the authors.

\bibliography{doe}

\end{document}